\documentclass[aps,prl,showpacs,twocolumn,amsmath,amssymb,superscriptaddress,footinbib]{revtex4}
\usepackage[english]{babel}
\usepackage{latexsym}
\usepackage{graphics}
\usepackage{subfigure}
\usepackage{epsfig}

\begin{document}

\title{Suppression of Transport of an Interacting Elongated Bose-Einstein Condensate in a Random Potential}
\author{D.~Cl\'{e}ment}
\author{A.F.~Var\'{o}n}
\author{M.~Hugbart}
\author{J.A.~Retter}
\author{P.~Bouyer}
\author{L.~Sanchez-Palencia}
\affiliation{Laboratoire Charles Fabry, Institut d'Optique, Universit\'e Paris-Sud XI,
91403 Orsay cedex, France}
\author{D.M.~Gangardt}
\affiliation{Laboratoire de Physique Th\'{e}orique et Mod\`{e}les
Statistiques, Universit\'e Paris-Sud XI, 91405 Orsay cedex, France}
\author{G.V.~Shlyapnikov}
\affiliation{Laboratoire de Physique Th\'{e}orique et Mod\`{e}les
Statistiques, Universit\'e Paris-Sud XI, 91405 Orsay cedex, France}
\affiliation{Van der Waals-Zeeman Institute, University of Amsterdam, Valckenierstraat 65/67, 1018 XE Amsterdam, The Netherlands}
\author{A.~Aspect}
\affiliation{Laboratoire Charles Fabry, Institut d'Optique, Universit\'e Paris-Sud XI,
91403 Orsay cedex, France}

\date{\today}

\begin{abstract}
We observe the suppression of the 1D transport of an interacting
elongated Bose-Einstein condensate in a random potential with a
standard deviation small compared to the typical energy per atom,
dominated by the interaction energy. Numerical solutions of the
Gross-Pitaevskii equation reproduce well our observations. 
We propose a scenario for disorder-induced trapping of the condensate in agreement with our observations.
\end{abstract}


\maketitle

Atomic Bose-Einstein condensates (BEC) in optical potentials
are a remarkable system in which to revisit standard problems of
condensed matter physics, 
{\it e.g.} superfluidity and quantum
vortices, the superfluid to Mott insulator transition, 
or Josephson arrays \cite{natureinsights}. Another
important topic in condensed matter physics is that of transport
in disordered materials, with relevance to
normal metallic conduction,
superconductivity and superfluid flow in low temperature quantum
liquids. 
This is a difficult
problem and it has led to the introduction of intriguing
and non-intuitive concepts, {\it e.g.} Anderson
localization \cite{anderson,andersonbis}, percolation 
\cite{aharony} and Bose \cite{fisher} and spin \cite{parisi}
glasses. It also has a counterpart in wave physics, {\it e.g.} in
optics and acoustics, specifically coherent diffusion in
random media \cite{Akkerman}. The main difficulty in
understanding quantum transport arises from the subtle interplay
of interference, scattering onto the potential landscape, and
(whenever present) interparticle interactions.

Transport properties of BECs in periodic optical lattices have been
widely investigated, showing lattice-induced reduction of mobility
\cite{PRL86,PRL88,PRL94} and self-trapping \cite{Oberthaler}.
Within the context of random potentials, most of the recent
theoretical efforts \cite{ahufinger2005}
have considered disordered or quasi-disordered
optical lattices where a large variety of phenomena have been
discussed such as the Bose-glass phase transition
\cite{damski2003}, localization \cite{damski2003,quasi2005}, and
the formation of Fermi-glass, quantum percolating and spin-glass phases
in Fermi-Bose mixtures \cite{spinglass,ahufinger2005}. 
Effects of disorder on BECs
have also been addressed in connection to superfluid flows in liquid helium
in porous media \cite{glyde2000}.
In particular, 
the depletions of the condensate
and of the superfluid fractions have been calculated
in Ref.~\cite{huang1992}, and a significant shift and damping of
sound waves have been predicted in Ref.~\cite{giorgini1994}. Apart from the (undesired) fragmentation effect
of a rough potential on trapped cold atoms and BECs on atom chips
\cite{chips}, there are few experiments on BECs in random
potentials \cite{massimo}.

\begin{figure}[b!]
\begin{center}
\includegraphics[width=7.cm]{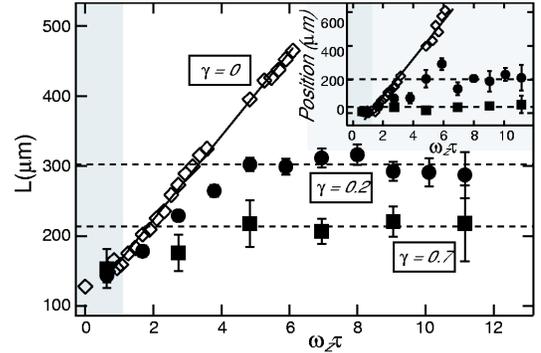}
\end{center}
\caption{Time evolution of the axial rms size 
$L$ of the BEC,
for various amplitudes $\sigma_V$ of the random potential, 
all smaller than the chemical potential  $\mu$
[$\gamma=\sigma_V/\mu = 0$ ($\diamond$), 
$0.2$ ($\bullet$) and 
$0.7$ ($\blacksquare$)].
The axial trapping frequency is initially $\omega_z / 2\pi = 6.7$~Hz
and is relaxed during the
first $30$~ms ($\omega_z \tau < 1.26$) of the expansion time (grey band).
Each point corresponds to an average over three measurements; 
error bars represent one standard deviation. 
The solid lines are linear fits to the data and the dashed lines are guides to the eye.
Inset: Motion of
the center of mass of the BEC during axial expansion for the
same values of $\gamma$.
Both sets of data show a strong suppression of transport of the BEC in the presence
of disorder.} 
\label{expansion}
\end{figure}

In this Letter we report on the strong reduction of mobility of
atoms in an elongated BEC in a random
potential~\cite{inguscio}. Starting from a  
BEC in a 3D highly elongated harmonic trap, we turn off the axial
trapping potential while maintaining strong transverse
confinement, and we monitor both (i) the axial expansion driven by the
repulsive interactions and (ii) the motion of the center
of mass of the BEC.
When the BEC is subjected
to a 1D random potential created by
laser speckle, the axial expansion is strongly inhibited and the BEC 
eventually stops expanding (see Fig.~\ref{expansion}). The final 
rms size $L$
decreases as the standard deviation $\sigma_{V}$ of the random
potential increases. 
The same effect has been observed for various
realizations of the random potential. We also observe that the
center of mass motion provoked by a longitudinal magnetic `kick'
at the time of release is strongly damped and is stopped in about
the same time (see Fig.\,\ref{expansion}). These observations are
\emph{not} made in a regime of tight binding, \emph{i.e.}~we
observe this localization effect
\footnote{We use the word `localization' in its `working definition' 
of absence of diffusion, see 
B.~van~Tiggelen, in {\it Wave Diffusion in Complex Media},
lectures notes at Les Houches 1998, edited by J.~P.~Fouque, NATO
Science (Kluwer, Dordrecht, 1999).} 
for amplitudes of the random potential 
small compared to the chemical potential. 
One may wonder whether our observations can be interpreted in terms of  
Anderson localization \cite{anderson}. In fact, in our situation, the interaction 
energy plays a crucial role, and the healing length is smaller than 
the typical distance between the speckle grains. 
This implies a different scenario, which we discuss in
this Letter.

We create an elongated $^{87}$Rb BEC in an iron-core
electromagnet Ioffe-Pritchard trap \cite{orsayBEC,Bragg} with
oscillation frequencies, 
$\omega_{\bot} / 2\pi = 660(4)\,$Hz
radially and  $\omega_{z} / 2\pi = 6.70(7)\,$Hz axially.
BECs of typically $3.5\times 10^5$ atoms are obtained, with
Thomas-Fermi half-length 
$L_{\rm{TF}}=150\,\mu$m and radius
$R_{\rm{TF}}=1.5\,\mu$m, and chemical potential 
$\mu/2\pi\hbar \sim 5\,$kHz
\footnote{
In elongated traps, thermal phase fluctuations may be
important for large enough temperatures \cite{Pet01}. 
However, here, we estimate the temperature to be
$150\,$nK \cite{Ger04} and therefore the phase coherence length
\cite{Bragg} is $L_{\phi}\sim L_{\rm{TF}}$, implying that the
BEC is almost fully coherent.}.   
The random potential is turned on at the
end of the evaporative cooling ramp and we further 
evaporate during 200\,ms to ensure that the BEC is in equilibrium
in the combined harmonic plus random potential at the end of 
the sequence.

\begin{figure}[t!]
\begin{center}
\includegraphics[width=6.5cm]{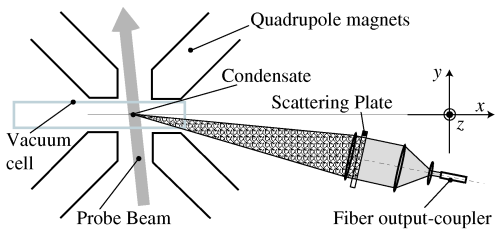}
\end{center}
\caption{Optical setup used to create the random speckle
potential. 
The BEC is at the focus of the lens system with its long axis
oriented along the $z$ direction.}\label{setup}
\end{figure}

To create the random potential, a $P\leq 150\,$mW blue detuned laser 
beam with optical wavelength $\lambda \simeq 780\,$nm,
perpendicular to axis $z$ is shone through a scattering plate 
and projects a speckle pattern \cite{goodman}
on the BEC (see Fig.~\ref{setup}). 
The scattered beam diverges to an rms radius of
1.83\,mm at the BEC.

A speckle field is defined by 
(i) a random intensity $I(\bf{r})$ with exponential statistical 
distribution for which  
the standard deviation equals the average intensity
$\sigma_I=\langle I \rangle$ 
and (ii) an intensity correlation
length $\Delta z$, defined as the
`half-width' of the autocorrelation function \cite{goodman}:
\begin{equation}
\Delta z= 1.22\ \lambda l/D, \label{Aofu}
\end{equation}
where $D$ is the beam diameter at the scattering plate and $l$ is
the distance from the lens to the BEC.
We observe the speckle intensity distribution on a
CCD camera placed at the same distance as the atoms. 
From this, we determine the autocorrelation
function to obtain the grain size $\Delta z$
for various beam diameters $D$. 
Taking into account
the modulation transfer function \cite{Hug05} of the camera,
we find that the measured grain size obeys Eq.~(\ref{Aofu}) to
within 2\%.
For our setup ($l=140(5)\,$mm and $D=25.4(1)\,$mm), Eq.~(\ref{Aofu}) gives
$\Delta z= 5.2(2)~\mu$m. This is an order of magnitude greater
than the healing length $\xi = (8\pi n a)^{-1/2}= 0.11~\mu$m of
the trapped BEC. 
Since $R_{\rm{TF}} < \Delta z \ll L_\textrm{TF}$,
the optical potential is effectively 1D,
with the trapped BEC spread over about $45-50$ wells in the
axial direction. We characterize the amplitude of the
random potential $\sigma_V$ with respect to the chemical potential
$\mu$ by \cite{Grimm}:
\begin{equation}
\gamma = \frac{\sigma_V}{\mu}=\frac{2}{3}\frac{\Gamma^2}{
2\delta}\frac{\sigma_{I}}{I_{\rm{S}}}\frac{1}{\mu}
=\frac{1}{\overline{\omega}}\frac{\Gamma^2}{6
\delta}\frac{\sigma_{I}}{I_{\rm{S}}}\left(\frac{15 a
N}{\overline{a_{\rm{ho}}}}\right)^{-2/5}
\end{equation}
with $\overline{\omega}$=$(\omega_{z} \omega_{\bot}^{2})^{1/3}$
and $\overline{a_{\rm{ho}}}$=$(\hbar/m \overline{\omega})^{1/2}$,
$m$ the atomic mass, $N$ the BEC atom number,
$I_{\rm{S}}=16.56$\,W/m$^2$ the saturation intensity,
$\Gamma/2\pi=6.01\,$MHz the linewidth, $a=5.31\,$nm the
scattering length and $\delta$ the laser detuning (between $0.15$~nm and 
$0.39$~nm in wavelength).
The factor $2/3$ accounts for the
transition strength for $\pi$-polarized light. Taking into account our
calibration uncertainty, we 
measure $\gamma$ within $\pm 20\%$.
For our parameters, the spontaneous scattering
time $1/\Gamma_\textrm{sc}$ 
is always larger than $1$s, {\it i.e.}
much longer than the experiment.

To study the coherent transport of the BEC in
the random potential, we open the axial magnetic trap
while keeping the transverse confinement and the random
potential unchanged.  
After lowering the current in the axial excitation coils,
the axial trapping frequency $\omega_{z} / 2\pi$ is 
smaller than $1$~Hz \footnote{From the
variations of the dipole curvature with the current, we
estimate the upper bound of $\omega_{z}/2\pi$ to be 
$\sim 500$\,mHz, which is compatible
with the linear expansion observed in the absence of disorder (see Fig.~\ref{expansion}).}.
Opening the trap abruptly induces atom loss and heating,
therefore the trap is opened in 30\,ms to avoid these processes. 
Once the current in the axial
coils has reached its final value we have a BEC of $N \sim
2.5\times 10^5 - 3\times 10^5$ atoms in the magnetic guide.

After a total axial
expansion time $\tau$ (which includes the 30~ms opening time), 
we turn off all remaining fields
(including the random potential) and wait a further 15\,ms of free
fall before imaging the atoms by absorption. 
During this
time-of-flight, the axial rms size of the BEC does not increase more than $5\%$. 
From profiles of the absorption images we evaluate the axial 
rms size $L$ 
\footnote{The rms axial size is defined
$L=\sqrt{\langle (z -\langle z\rangle)^2\rangle}$
where $\langle \cdot \rangle$ stands for the density-weighted average.}
which we plot in Fig.~\ref{expansion} versus the axial expansion time $\tau$. 
In the absence of
the random potential ($\gamma=0$), we observe that the 
rms size $L$ 
grows linearly at a rate
$v_{\rm RMS} \sim 2.47(3)$\,mm\,s$^{-1}$
in agreement with the scaling theory \cite{scaling}. 
In the presence of the random
potential, the expansion dynamics changes dramatically. For a
sufficiently high amplitude, the expansion
is significantly reduced and the BEC eventually stops
expanding. In addition, we observe the damping of longitudinal
motion of the center of mass of the BEC (see inset of
Fig.\,\ref{expansion}). This motion is triggered by an axial
magnetic `kick' during the opening of the trap.

\begin{figure}[t!]
\begin{center}
\includegraphics[width=6.cm]{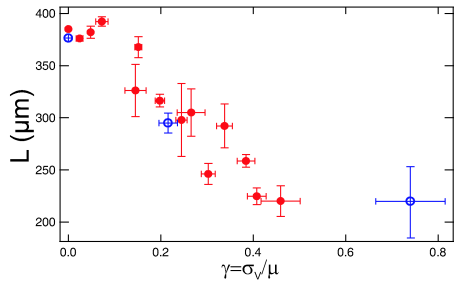}
\end{center}
\caption{Rms size $L$ of the BEC versus $\gamma$ after
an axial expansion time $\omega_z \tau=4.84$ ($\tau=115$ms). 
The open circles correspond to the curves of Fig.~\ref{expansion}.}
\label{l_power}
\end{figure}

These results show a transition from non-inhibited to inhibited 
transport as the speckle amplitude
is increased.
This is studied
in further detail by measuring the BEC rms size
after a fixed axial expansion time of
115~ms ($\omega_z \tau =4.84$) for different amplitudes $\sigma_V$ of the
random potential.  The results are shown in Fig.\,\ref{l_power}.
We see that above a value $\gamma=0.15$, the 
rms size
decreases with $\gamma$.

From the absorption images, we also evaluate the density in the
magnetic guide, after correcting for radial expansion during
the time-of-flight.  We observe that the density at the center of the
BEC does not drop by more than a factor of 2 for $\gamma > 0.2$.
Therefore, we conclude that the interaction energy dominates at
the center of the BEC trapped by disorder, a point we shall
discuss below.

To understand the suppression of expansion of the BEC in the random
potential, we have performed numerical calculations of the BEC
dynamics in the mean-field Gross-Pitaevskii approach. We
consider a BEC trapped in a cylindrically-symmetric 3D-harmonic
trap with frequencies $\omega_\bot$ and $\omega_z$ in
the radial and axial directions respectively.
Assuming tight radial confinement
($\hbar \omega_\bot \gg \hbar \omega_z, \mu, k_\textrm{B}T$), the 
dynamics is reduced to 1D. 
In addition, the BEC is subjected to a static random potential 
$V(z)=\sigma_{V} v(z)$ where
$v(z)$ is a normalized numerically-generated speckle pattern \cite{goodman} 
with $\langle v \rangle^2 = \langle v^2 \rangle /2 = 1$.
This slightly differs from the experimental situation where the BEC
is very elongated but not strictly 1D. 
However, in the experiment, the BEC is {\it guided} in a 1D random potential 
so that the radial size only slightly changes and,
due to the different time scales in the axial ($1/\omega_z$) and radial
($1/\omega_\bot \ll 1/\omega_z$) directions, the radial size adapts 
adiabatically to the axial size.
Thus we expect that the 1D simplified model captures the physics of the
experiment.
We consider parameters close to the experimental situation (see above). 
In particular, the healing length ($\xi \simeq 8\times 10^{-4} L_\textrm{TF}$) 
and the speckle correlation length ($\Delta z \simeq 0.049 L_\textrm{TF}$)
are much smaller than the 
size of the BEC.

\begin{figure}[t!]
\begin{center}
\includegraphics[width=8.5cm]{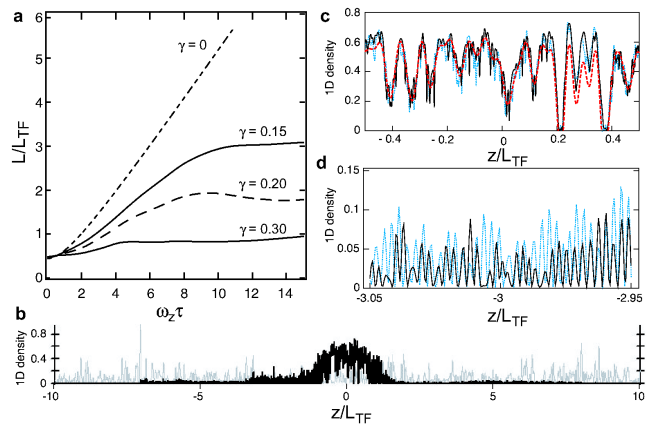}
\end{center}
\caption{a) Time evolution of the rms size $L$ of the BEC in the random potential 
$V(z)$ for various values of the speckle amplitude $\sigma_V=\gamma \mu$ as 
obtained from the numerical calculations. 
b) Density profile (black) and random potential $V(z)/g_\textrm{1D}$ (gray) 
for $\gamma = 0.2$ at $\omega_z \tau=10$.  
c-d) Enlargement of density profile at $\omega_z \tau=10$ (black solid) and $\omega_z \tau=20$ (blue dotted).  
The red dashed line in c) is the Thomas-Fermi prediction (see text).}
\label{expan}
\end{figure}

We first compute the static 1D BEC wavefunction in the combined (harmonic plus random) trap.
Because $\xi \ll \Delta z$, the density profile simply follows
the modulations of the combined trap in
the Thomas-Fermi regime:
$|\psi (z)|^2 = [\mu -m\omega_z^2 z^2/2-V(z)]/g_\textrm{1D}$
in the region where $\mu > m\omega_z^2 z^2/2 + V(z)$ and $|\psi (z)|^2=0$ elsewhere.
Here, $m$ is the atomic mass and 
$g_\textrm{1D}=2\hbar a \omega_\bot$ the 1D interaction 
parameter.
At time $\tau=0$, we suddenly switch
off the axial harmonic confinement
while keeping
unchanged the interaction parameter $g_\textrm{1D}$ and the random
potential and we compute the time-evolution of the BEC. The
results for the axial rms size
$L$ of the BEC are plotted in
Fig.~\ref{expan}a for various amplitudes of the random potential.
In the absence of disorder, the evolution of the BEC corresponds
to self-similar expansion 
with scaling parameter $b(t) \sim \sqrt{2} \omega_z t$ 
\footnote{Our findings are in excellent agreement with the analytic
solution of the scaling equations of \cite{scaling},\\
$\sqrt{b(t) [b(t)-1]} + \ln [\sqrt{b(t)}+\sqrt{b(t)-1}] = \sqrt{2} \omega_z t$.}.
In the presence of disorder ($\gamma \gtrsim 0.15$), after
initial expansion, the BEC stops expanding.
This is qualitatively the same behavior we observed in the 
experiment. 
The quantitative agreement is also reasonably good. 
For example, for $\gamma=0.2$, the BEC expands by a factor
of $\simeq 4$ in the numerics ($\simeq 3$ in the experiment)
and is trapped after a transient expansion time of 
$\omega_z \tau \simeq 8$ ($\omega_z \tau \simeq 6$).
This strong suppression of expansion corresponds to
disorder-induced trapping of the BEC.

We now describe a scenario for disorder-induced trapping of the BEC
after release of the axial harmonic confinement. 
For small enough amplitudes of the random potential, the initial stage of expansion 
can be described using the scaling theory \cite{scaling}. 
According to this, the fast atoms populate the
wings of the expanding BEC whereas the slow atoms are close to the center.
It is thus tempting to distinguish two regions of
the BEC:
(i) the center where the interaction energy dominates the 
kinetic energy and trapping is due to
the competition between interactions and disorder, and
(ii) the wings where the 
kinetic energy exceeds the interactions
and trapping is rather due to the competition between
the kinetic energy and disorder.

In the center, the average density and thus the effective chemical
potential $\overline{\mu}$ slowly decrease during the expansion
stage. As the interaction energy is much larger than the
kinetic energy, the local density adiabatically follows the
instantaneous value of $\overline{\mu}$ in the Thomas-Fermi regime:
$|\psi (z)|^2 = [ \overline{\mu} - V(z) ] / g_\textrm{1D}$ in the region
where $\overline{\mu} > V(z)$ and $|\psi (z)|^2 = 0$ elsewhere.
This agrees with our numerical results (see Fig.~\ref{expan}c).
This evolution stops with fragmentation, 
{\it i.e.} when the BEC meets two peaks of the random potential
with amplitudes larger than $\overline{\mu}$.
Using the statistical properties of the random potential \cite{goodman},
we can estimate the probability of such large peaks and we conclude
that this happens when the central density $n_0$
reaches the value
\begin{equation}
n_0 \simeq 1.25 \left( \frac{\sigma_\textrm{V}}{g_\textrm{1D}} \right) \ln \left[ \frac{0.47 L_\textrm{TF}}{\Delta z} \right] ~.
\label{n0eq}
\end{equation}
This formula is in good agreement with our numerical and experimental findings
\cite{longpaper}.

Due to the small density, the situation is completely different in the
wings which are populated by almost free particles interacting
with the disordered potential. The BEC wavefunction thus undergoes 
disorder-induced multiple reflections and transmissions
and is ultimately blocked by a large peak of the speckle
potential. Therefore, the BEC is not in the Thomas-Fermi regime and
the local density is not stationary (see Fig.~\ref{expan}d). Due to conservation of
energy, the 
kinetic energy per particle  $\epsilon$ is of the
order of the typical energy in the initial BEC ($\epsilon
\sim \mu$) so that the typical wavelength
of the fluctuations in the wings is of the order of the healing length
in the initial BEC 
$\lambda_\textrm{w} \sim \xi = \hbar/\sqrt{2m\mu}$.

This scenario of disorder-induced trapping is accurately supported by our
numerical integration of the Gross-Pitaevskii equation. In particular,
the density profiles plotted in Fig.~\ref{expan} 
show the static Thomas-Fermi shape in the center and time-dependent
fluctuations in the wings with typical wavelength 
$\lambda_\textrm{w} \sim \xi$ \cite{longpaper}.

In conclusion, we have experimentally investigated transport
properties of an interacting BEC in a random potential.
Controlling the strength of disorder, we have observed the
transition from free expansion to absence of diffusion as disorder
increases. 
We have presented numerical simulations that reproduce well the
observed suppression of expansion and we have discussed a theoretical model that describes the
scenario for disorder-induced trapping. In the future, it would be
interesting to further investigate this highly controllable
system, for example by changing the correlation length of disorder
or employing Bragg spectroscopy to probe the momentum spectrum of
the BEC \cite{Bragg}.

We are grateful to P.~Chavel, J.~Taboury, F.~Gerbier
and C.~Henkel for fruitful discussions. 
We acknowledge
support from IXSEA-OCEANO (M.H.) and the Marie Curie Fellowships
programme of the European Union (J.R.). This work was supported by
CNRS, D\'{e}l\'{e}\-gation G\'{e}n\'{e}rale de l'Armement, 
Minist\`ere de la Recherche (ACI Nanoscience 201) and the
European Union (grants IST-2001-38863 and MRTN-CT-2003-505032) and
INTAS (Contract 211-855).




\begin{references}
\bibitem{inguscio}
While writing this paper, we have been informed that similar systems
are currently studied
in the groups of M.~Inguscio at
LENS and 
of W.~Ertmer in Hannover (private communications).
\bibitem{longpaper}
A detailed analysis of these properties will be described in a
future publication.
\end{references}
\end{document}